\documentclass[epj]{webofc}
\usepackage[utf8]{inputenc}
\usepackage[varg]{txfonts}   
\usepackage{booktabs}
\usepackage{xcolor}
\definecolor{darkred}{rgb}{0.4,0.0,0.0}
\definecolor{darkgreen}{rgb}{0.0,0.4,0.0}
\definecolor{darkblue}{rgb}{0.0,0.0,0.4}
\usepackage[bookmarks,linktocpage,colorlinks,
    linkcolor = darkred,
    urlcolor  = darkblue,
    citecolor = darkgreen]{hyperref}
\usepackage{subfigure}
\usepackage[flushleft]{threeparttable}
\wocname{EPJ Web of Conferences}
\woctitle{Lattice2017}
\def\proof{\noindent{\sl Proof:}\kern0.6em}

\def\frac#1#2{\hbox{$#1\over#2$}}
\def\dual{\mathstrut^*\kern-0.1em}

\def\lvec#1{\setbox0=\hbox{$#1$}
    \setbox1=\hbox{$\scriptstyle\leftarrow$}
    #1\kern-\wd0\smash{
    \raise\ht0\hbox{$\raise1pt\hbox{$\scriptstyle\leftarrow$}$}}
    \kern-\wd1\kern\wd0}
\def\rvec#1{\setbox0=\hbox{$#1$}
    \setbox1=\hbox{$\scriptstyle\rightarrow$}
    #1\kern-\wd0\smash{
    \raise\ht0\hbox{$\raise1pt\hbox{$\scriptstyle\rightarrow$}$}}
    \kern-\wd1\kern\wd0}
\def\slash#1{\setbox0=\hbox{$#1$}\setbox1=\hbox{$\kern1pt/$}
    #1\kern-\wd0\kern1pt/\kern-\wd1\kern\wd0}

\def\nab#1{{\nabla_{#1}}}
\def\nabstar#1{{\nabla\kern0.5pt\smash{\raise 4.5pt\hbox{$\ast$}}
               \kern-5.5pt_{#1}}}

\def\nabbar#1{{\overset{\leftarrow}{\nabla}_{#1}}}
\def\nabbarstar#1{{\overset{\leftarrow}{\nabla}\kern0.5pt\smash{\raise 4.5pt\hbox{$\ast$}}
               \kern-5.5pt_{#1}}}

\def\drvstar#1{{\partial\kern0.5pt\smash{\raise 4.5pt\hbox{$\ast$}}
               \kern-6.0pt_{#1}}}

\def\ldrvstar#1{{\lvec{\,\partial}\kern-0.5pt\smash{\raise 4.5pt\hbox{$\ast$}}
               \kern-5.0pt_{#1}}}

\def\MeV{{\rm MeV}}
\def\GeV{{\rm GeV}}

\def\fm{{\rm fm}}
\def\MSbar{\overline{\rm MS\kern-0.5pt}\kern0.5pt}
\def\Nf{{N_{\rm f}}}

\def\psibar{\overline{\psi}}

\def\zetabar{\bar{\zeta}}
\def\zetaprime{\zeta\kern1pt'}
\def\zetabarprime{\zetabar\kern1pt'}

\def\diracstar#1#2{
    \setbox0=\hbox{$\gamma$}\setbox1=\hbox{$\gamma_{#1}$}
    \gamma_{#1}\kern-\wd1\kern\wd0
    \smash{\raise4.5pt\hbox{$\scriptstyle#2$}}}

\def\SUthree{{\rm SU(3)}}

\def\Ds{D_{\rm s}}
\def\DsdagDs{\Ds{\Ds}^{\kern-1pt\dagger}}

\def\avg#1{{\kern1.0pt\overline{\kern-1.0pt#1\kern-1.0pt}\kern1.0pt}}

\begin{document}
%
\hfill CERN-TH-2017-220  

\selectlanguage{english}
\title{QCD in a moving frame: an exploratory study}
\author{%
\firstname{Mattia} \lastname{Dalla Brida}\inst{1,2}\fnsep\thanks{Speaker, \email{mattia.dalla.brida@desy.de}} \and
\firstname{Leonardo} \lastname{Giusti}\inst{1,2,3} \and 
\firstname{Michele} \lastname{Pepe}\inst{2}
}
\institute{%
Dipartimento di Fisica, Universit\`a di Milano-Bicocca,
Piazza della  Scienza 3, I-20126 Milano, Italy
\and
INFN, Sezione di Milano-Bicocca, 
Piazza della  Scienza 3, I-20126 Milano, Italy
\and
CERN, Theoretical Physics Department, 1211 Geneva 23, Switzerland
}
\abstract{%
  The framework of shifted boundary conditions has proven to be a very powerful
  tool for the non-perturbative investigation of thermal quantum field theories. 
  For instance, it has been successfully considered for the determination
  of the equation of state of $\SUthree$ Yang-Mills theory with high accuracy. 
  The set-up can be generalized to QCD and it is expected to lead to a similar 
  breakthrough. We present first results for QCD with three flavours of 
  non-perturbatively O($a$)-improved Wilson fermions and shifted boundary
  conditions.
}
\maketitle
\section{Introduction}\label{intro}

The understanding of a large variety of phenomena involving the strong interactions, 
ranging from the dynamics inside a nucleon star to the evolution of the early
Universe, crucially depends on the accurate knowledge of the equation of state (EoS)
of QCD. In particular, those extreme conditions are now being reproduced and 
investigated at heavy-ion colliders, where the EoS is an essential input for the analysis
of the data~\cite{Dainese:2016gch}.

First principles determinations of the EoS of QCD are on the other hand very
challenging, as one needs non-perturbative control of QCD over a wide range of
temperatures. Even at relatively high temperatures, indeed, standard perturbative
methods are characterized by very poor convergence and elaborated techniques
are necessary to improve convergence (see e.g.~\cite{Kajantie:2002wa,Blaizot:2003iq,
Andersen:2004fp}). Most importantly, due to the asymptotic nature of the perturbative 
expansion, it is difficult (if not impossible) to access the accuracy of the 
results within perturbation theory itself (even if apparent convergence is seen), 
unless one can compare with non-perturbative data over a wide range of temperatures. 
Therefore, lattice QCD is at present the only known framework that allows us to 
tackle this problem from first principles, in a fully systematic and predictive way.

Results for the EoS of QCD with $\Nf=2+1$ quark flavours at zero chemical potential 
have been recently obtained using well-established lattice techniques~\cite{Borsanyi:2013bia,
Bali:2014kia,Bazavov:2014pvz}: the so-called integral method~\cite{Boyd:1996bx}
and its variants. However, although many interesting results have been obtained
with these methods, they become computationally very demanding as the temperature
is increased, thus limiting in practice the accessible range of temperatures. 
Most calculations are in fact confined to $T\lesssim500\ \MeV$. Only very recently
results at higher temperatures, i.e. up to $2\,\GeV$, began to appear~\cite{Bazavov:2017dsy}, 
but reliable continuum extrapolations are still difficult at the highest temperatures.

For the very same reason of high computational cost, all state of the art 
determinations have been exclusively obtained in the framework of staggered
fermions. It is, of course, of crucial importance at this point to provide independent
results with also other discretizations, in order to increase our confidence 
in the continuum results obtained so far.

In the last few years, a big effort has been
invested into devising new methods to address this problem, and overcome the limitations
of the current state of the art techniques. In this contribution we consider, 
for the first time, the framework of \emph{shifted boundary conditions} 
(SBC)~\cite{Giusti:2010bb,Giusti:2011kt,Giusti:2012yj} applied to QCD. This set-up 
has been successfully applied to the case of $\SUthree$ Yang-Mills 
theory~\cite{Robaina:2013zmb,Giusti:2014ila,Umeda:2014ula,Giusti:2015daa,
Giusti:2016iqr}, leading above all to a determination of the EoS over two orders
of magnitude in the temperature with half a per-cent accuracy ~\cite{Giusti:2014ila,Giusti:2015daa,
Giusti:2016iqr}. A similar breakthrough is expected in the case of QCD. 

In the following, the feasibility of a determination of the EoS in QCD with $\Nf=2+1$
non-perturbatively O($a$)-improved Wilson fermions and SBC is investigated, and it 
showed that precise results are possible over a wide range of temperatures.
Two orders of magnitude in the temperature, $0.36\,\GeV\lesssim T\lesssim 45\,\GeV$,
could easily be covered.

\section{QCD in a moving frame} \label{sec-2}

In a relativistic thermal field theory one can relate the entropy of the 
system in its local rest frame to the momentum measured by an observer 
in a moving reference frame (see e.g.~\cite{Landau:1982dva}). Hence, if 
one is able to measure the momentum of the system in a moving frame, one
can directly access its entropy in its rest frame. Any other thermodynamic
potential can be obtained from standard thermodynamic relations. 

Remarkably, the Euclidean path integral formulation of a thermal field 
theory in a moving frame is rather simple: it amounts to imposing shifted 
boundary conditions to the fields in the compact (time) direction~\cite{Giusti:2012yj}.
In the case of lattice QCD this means requiring:
\begin{gather}
 \label{eq:SBC}
 \nonumber
 U_\mu(L_0,\boldsymbol{x})=U_\mu(0,{\boldsymbol{x}}-{L_0}{\boldsymbol{\xi}}),\\
 \psi(L_0,\boldsymbol{x})=-\psi(0,{\boldsymbol{x}}-{L_0}{\boldsymbol{\xi}}),
 \quad
 \psibar(L_0,\boldsymbol{x})=-\psibar(0,{\boldsymbol{x}}-{L_0}{\boldsymbol{\xi}}),
\end{gather}
where $L_0$ is the physical extent of the compact direction, and ${\boldsymbol{\xi}}$
is the shift vector corresponding to the imaginary velocity of the system; periodic 
boundary conditions are assumed in the three spatial directions of extent $L$.

The entropy density $s(T)$ at the temperature $T$ can now be computed 
as~\cite{Giusti:2012yj}:
\begin{equation}
 \label{eq:Entropy}
 {s(T)\over T^3}=-{L_0^4(1+{\boldsymbol{\xi}^2})^3\over \xi_k}\,
 \langle T^R_{0k} \rangle_\xi,
 \quad
 \xi_k\neq0,
 \quad
 T^{-1}=L_0\sqrt{1+\boldsymbol{\xi}^2},
\end{equation}
where $\langle\cdots\rangle_\xi$ denotes the path-integral expectation value
in presence of SBC, and $T_{0k}^R$ corresponds to the momentum $k$-component
of the renormalized \emph{energy-momentum tensor} (EMT).%

From the results of the analysis of~\cite{Caracciolo:1989pt,Caracciolo:1991vc,
Caracciolo:1991cp}, the renormalized momentum components of the EMT can be written
as,
\begin{equation}
  T_{0k}^R(x)=Z_{T^F}(g_0)T^F_{0k}(x) +  Z_{T^G}(g_0)\, T^G_{0k}(x),
\end{equation}
where $T^F_{0k}$ and  $T^G_{0k}$ are the bare fermionic and gluonic components
of the EMT; $Z_{T^F}$ and $Z_{T^G}$ are two finite (i.e. renormalization scale
independent) renormalization constants, which depend on the exact discretization
of the QCD action and of the bare fields $T^F_{0k}$ and $T^G_{0k}$.

For the gluonic component of the EMT, we choose (see~\cite{Caracciolo:1989pt,
Giusti:2015daa} for a precise definition),
\begin{equation}
 \label{eq:TG}
 T^G_{0k}(x)= 
 {1\over g_0^2}\sum_{\alpha=0}^3\sum_{a=1}^8 F^a_{0\alpha}(x) F^a_{k\alpha}(x),
\end{equation}
where $g_0$ is the bare coupling and $F^a_{\mu\nu}$ are the color 
components of the usual clover discretization of the field strength tensor.

For the fermionic part, we take,
\begin{equation}
\begin{split}
  \label{eq:TF}
  T^F_{0k}(x)=
  \frac{1}{8}\Big\{
  &\psibar(x)\gamma_0\big[\nabstar k+\nab k]\psi(x)
  -\psibar(x)[\nabbarstar k+\nabbar k]\gamma_0\psi(x)\\
  +&\psibar(x)\gamma_k\big[\nabstar 0+\nab 0]\psi(x)
  -\psibar(x)[\nabbarstar 0+\nabbar 0]\gamma_k\psi(x)\Big\},\\
\end{split}
\end{equation}
where $\nab\mu,\nabstar\mu$ and $\nabbar\mu,\nabbarstar\mu$ are the usual
forward and backward covariant lattice derivatives, acting on either the
quark or anti-quark fields, respectively. (Once again, we refer the reader 
to~\cite{Caracciolo:1989pt} for a precise definition.) We note that in
this exploratory investigation we may neglect any O($a$) operator 
counterterm to these fields.

Finally, as anticipated, for the lattice action we consider $\Nf=2+1$ 
non-perturbatively O($a$)-improved Wilson fermions and, if not stated 
otherwise, the Wilson (plaquette) gauge action.

\section{The entropy in free-field theory}

The information from lattice perturbation theory can give one useful 
insight into the lattice artifacts of the theory, and allows 
educated guesses on which set of kinematic parameters might lead 
to small discretization errors in non-perturbative investigations.
Hence, it is useful to first study Eq.~(\ref{eq:Entropy}) in the 
limit of free quarks and gluons i.e. at the lowest order in perturbation 
theory. Of course, the information inferred from perturbation theory
is expected to be more accurate the higher is the temperature, and the
closer one is to the continuum limit. In particular, as the results are
expected to be more trustworthy at high temperatures where quark mass
effects are strongly suppressed, below we shall focus on the case of 
massless quarks. 

\begin{figure}[htb]
  \centering
  \includegraphics[scale=0.75]{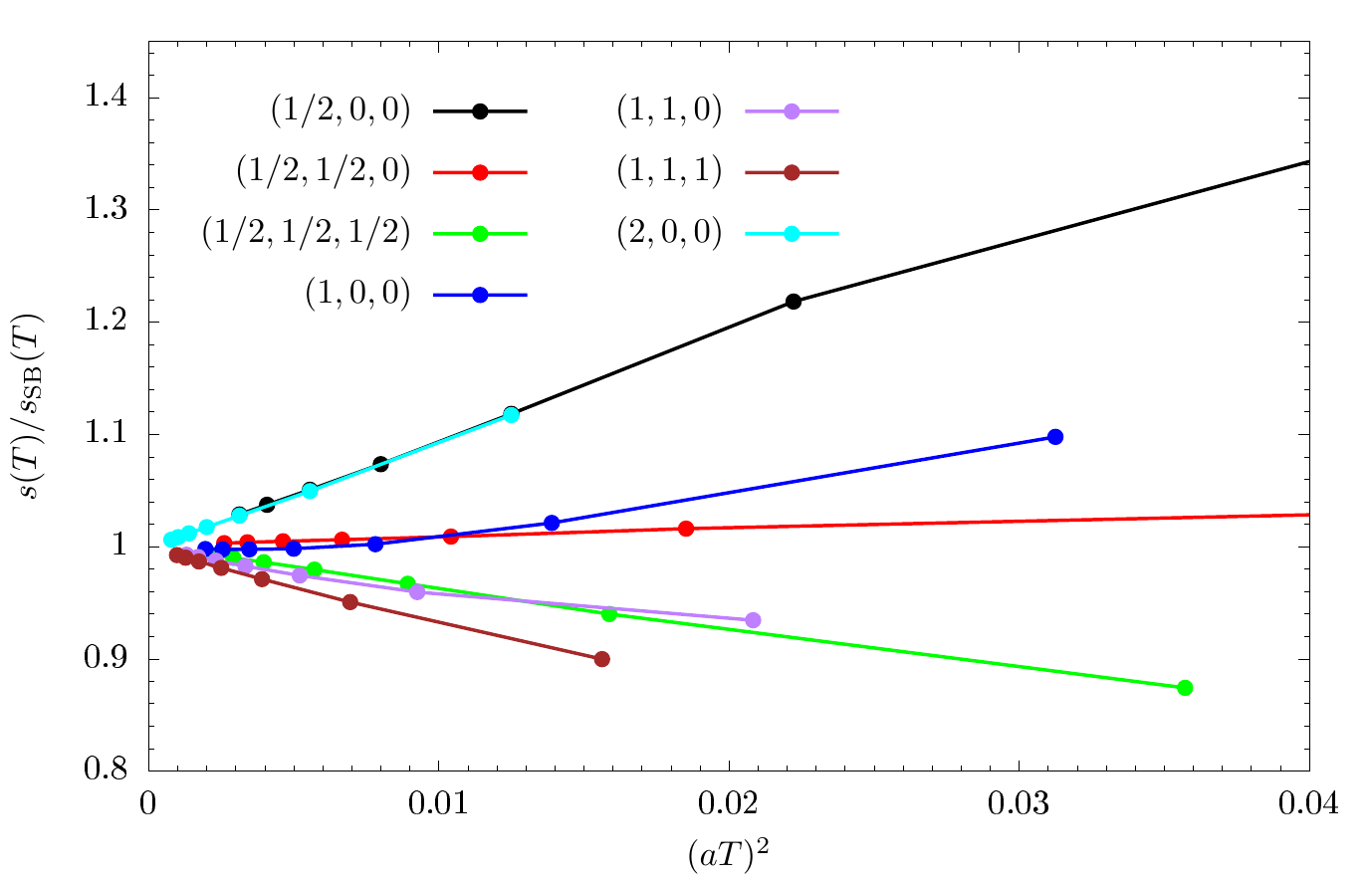}
  \caption{Continuum limit of the entropy density computed at the lowest-order 
	  in lattice perturbation theory normalized to the Stefan-Boltzmann 
	  limit, Eq.~(\ref{eq:SB}). The results are for $\Nf=3$, and different
	  choices of shift vector $\boldsymbol{\xi}$ are shown.}  
  \label{fig:LatticeEffects}
\end{figure}

In Figure \ref{fig:LatticeEffects} we show the continuum limit of the 
ratio of the entropy density computed at the lowest order in lattice 
perturbation theory through Eq.~(\ref{eq:Entropy}), and the expected
continuum limit, namely the Stefan-Boltzmann (SB) result for QCD
with $\Nf$ quark-flavours:
\begin{equation}
 \label{eq:SB}
 {s_{\rm SB}(T)\over T^3}= {\pi^2\over45}\big(32 + 21\times N_{\rm f}\big).
\end{equation}
Specifically, we present data for the relevant case of $\Nf=3$,
and different choices of shift vector $\boldsymbol{\xi}$.
We considered lattices with $L_0/a=4,\ldots,16$, and a fixed ratio 
$L/L_0=32$; with this choice finite volume effects can be ignored in
the present discussion. The results are plotted against the temperature
$T$ in lattice units (cf. Eq.~(\ref{eq:Entropy})). It is interesting
to note that, in the free case and for $\Nf=3$, the fermionic contribution
to the entropy density is twice as larger than the gluonic one. In particular,
not only the physical component of the entropy is dominated by the quark 
sector, but also its lattice artifacts. Nonetheless, the cases with 
$\boldsymbol{\xi}=(1,0,0)$ and $(1/2,1/2,0)$ appear to have in general very 
small lattice artifacts: below $5\%$ for $(aT)^2<0.02$. For these shift vectors,
the latter condition translates into having lattices with $L_0/a\geq 6$, for 
which discretization errors are about 2\% or below. Moreover, we note that the
approach to the continuum limit is clearly O($a^2$). This is expected since, 
in perturbation theory, the O($a$) effects in this quantity are all of 
O($am_q$), where $m_q$ is the subtracted bare quark mass. These effects are 
hence absent in the chiral limit.

\section{Simulation results}

The numerical simulations have been performed using a customized version of 
the \texttt{openQCD-1.6} package~\cite{Luscher:2012av,LuscherWeb:2016}.
This allowed us to employ several efficient algorithms to speed up the 
simulations. More precisely, the simulation of the doublet of up and down
quarks employed an optimized twisted-mass Hasenbusch preconditioning of the
quark determinant~\cite{Hasenbusch:2001ne}, while the strange quark was 
simulated through a RHMC algorithm~\cite{Kennedy:1998cu,Clark:2003na} with an
optimized frequency splitting of the rational approximation. Even-odd 
preconditioning was used for both the light and strange quarks. The integration
of the molecular dynamics equations was based on a three-level integration scheme. 
The gauge force was integrated on the finest level using a 4th-order Omelyan-Mryglod-Folk
(OMF4) integrator~\cite{Omelyan:2013}, while the fermionic forces were integrated
on the two coarser levels. On the finest of these we used a OMF4 integrator, while 
on the coarsest a 2nd-order OMF integrator~\cite{Omelyan:2013}. The solution of the
Dirac equation along the molecular dynamics evolution was obtained using a standard
conjugate gradient with chronological inversion. More details on the exact implementation
of these algorithms can be found in the references provided.

In order to access the feasibility of a determination of the EoS in 
3-flavour QCD with Wilson fermions and SBC, the first crucial issue to be
addressed is the computational effort necessary to obtain a given statistical
precision for the basic \emph{bare} quantities $\langle T^{F,G}_{0k} \rangle_\xi$;
particularly important is how this depends on the temperature. We considered 
three different ensembles of well separated temperatures; a detailed description is
given below. For all simulations we chose $\boldsymbol{\xi}=(1,0,0)$, $L_0/a=6$, and
$L/a=96$, therefore having $TL\approx11$. We note that the leading finite volume effects
in the entropy density are exponentially small in the product $ML$, where $M$ is the
lightest screening mass at the given temperature (see~\cite{Giusti:2012yj} 
for explicit formulas in presence of SBC). Having this noticed, based on the results
of finite volume studies conducted in $\SUthree$ Yang-Mills theory~\cite{Giusti:2014ila,
Giusti:2016iqr}, and on perturbative estimates of the lightest screening masses
in QCD~\cite{Laine:2009dh}, we expect finite volume effects to be well below the
precision reached in our exploratory runs (see below). Of course, for a precise 
determination  of the entropy density a systematic study of finite volume effects
needs to be carried out. In this respect, we emphasize that considering larger 
spatial volumes is not a real issue in our set-up. Our computation of
the entropy density is based on the measurement of a simple one-point function
of the EMT (cf. Eq.~(\ref{eq:Entropy})). The increase in computational effort due to
a larger spatial volume is hence largely compensated by the statistical gain deriving
from averaging the one-point function over a larger number of space-time points. Also,
simulating large spatial volumes would help in controlling systematic effects 
arising from the freezing of topology at small lattice spacings~\cite{Brower:2003yx,
Luscher:2017cjh}.

\begin{table}[hptb]
  \centering
  \begin{threeparttable}
  \begin{tabular}{lllll}
  \toprule
  Ensemble & $\beta$ & $c_{\rm sw}$ & $\kappa_{ud} $ & $\kappa_s$ \\
  \midrule
   $T_1{}^\dag$ & 3.5500 & 1.824865382 & 0.137080000 & 0.136840284 \\
   $T_2$        & 6.4680 & 1.409845309 & 0.135201231 & 0.135201231 \\
   $T_3$        & 8.0891 & 1.255925917 & 0.132802475 & 0.132802475 \\
  \bottomrule
  \end{tabular}
  \begin{tablenotes}
  \footnotesize 
  \item${}^\dag$ This ensemble employs the L\"uscher-Weisz gauge action instead 
  of Wilson's plaquette.
  \end{tablenotes}
  \end{threeparttable}
  \caption{Bare action parameters for the simulated ensembles at temperatures:
  $T_1$, $T_2$, $T_3$. As usual, $\beta=6/g_0^2$, and $\kappa^{-1}=2m_0+8$, 
  with $m_0$ the bare quark mass of either the up and down quarks ($ud$) or 
  the strange quark ($s$). $c_{\rm sw}$ denotes the improvement coefficient
  of the Sheikholeslami-Wohler term.}
  \label{tab:LatticePrams}
\end{table}

We shall now describe in some detail the three ensembles we considered.

\begin{enumerate}

 \item $\boldsymbol{T_1}$: The lattice bulk-action and action bare parameters
 of this ensemble match those of ensemble N203 of CLS~\cite{Bruno:2014jqa, Bruno:2016plf}. 
 Specifically, the simulations employ $\Nf=2+1$ non-perturbatively O($a$)-improved
 Wilson fermions and the Symanzik tree-level O($a^2$)-improved (L\"uscher-Weisz)
 gauge action (see the given references for more details). The bare action parameters
 are reported in Table \ref{tab:LatticePrams} for the reader's convenience. These 
 correspond to having a lattice spacing $a\approx 0.064\ \fm$, and quark masses
 such that $m_\pi\approx340\,\MeV$ and $m_K\approx440\,\MeV$ at $T=0$. The resulting
 temperature with our choice $L_0/a=6$ and $\boldsymbol{\xi}=(1,0,0)$ is 
 $T_1\approx 0.36\ \GeV$.
 
 \item $\boldsymbol{T_2}$: For this ensemble the bare coupling was fixed 
 by requiring the Sch\"odinger functional (SF) coupling at the energy scale 
 $\mu_0=T_2\times3\sqrt{2}/4$ to have the value $\bar{g}^2_{\rm SF}(\mu_0)=
 2.012$~\cite{Brida:2016flw,DallaBrida:2016kgh}. From the results 
 of~\cite{Bruno:2017gxd} one infers that $T_2\approx 4\ \GeV$. The bare quark
 masses were all set to approximatively their critical value, hence simulating
 the theory at very small quark masses. This was feasible since at high 
 temperature the massless Dirac operator develops a spectral gap $\lambda\propto T$.%
	\footnote{At lowest order in perturbation theory $\lambda=\pi T$.}
 The critical mass was estimated by extrapolating to infinite volume
 the results obtained in finite volume SF simulations~\cite{DallaBrida:2016kgh}.%
	\footnote{In principle there is no need to extrapolate the results
	for the critical mass determined in small volume SF simulations to infinite
	volume, as the volume dependence is a pure lattice artifacts. By performing
	an infinite volume extrapolation, however, we aimed at reducing such lattice
	effects.} 
 The precise set of parameters considered is given in Table \ref{tab:LatticePrams}.  
 In this respect, we must note that using the results for the quark-mass 
 renormalization determined along the lines of constant physics defined through the
 SF couplings~\cite{Campos:2016eef}, it will be possible in future runs to
 tune the quark masses to the their physical values over a wide range of 
 temperatures, i.e. up to $T\approx 70\,\GeV$.%
	\footnote{In fact, as quark mass effects are expected to be strongly suppressed
	at high temperature, we expect that a precise tuning of the quark masses will not
	be necessary in practice for a significant part of this temperature range.}
  
 \item $\boldsymbol{T_3}$: Analogously to the case of the $T_2$ ensemble,
 also for this ensemble we fixed the bare coupling through the SF coupling, 
 having $\bar{g}^2_{\rm SF}(\mu_{\rm})\approx 1.27$ with 
 $\mu=T_3\times3\sqrt{2}/4$. This corresponds to $T_3\approx45\ 
 \GeV$~\cite{Brida:2016flw,Bruno:2017gxd}. The quark masses were all set 
 (approximatively) to their critical value, estimated from small volume 
 SF simulations (cf. Table \ref{tab:LatticePrams}).
 
\end{enumerate}

\begin{table}[hptb]
  \centering
  \begin{threeparttable}
  \begin{tabular}{lrrrr}
  \toprule
  $T\,(\GeV)$ & $\langle T_{0k}^F\rangle_\xi\times10^4$ & $\langle T_{0k}^G\rangle_\xi\times10^4$ & 
  $N_{\rm ms}$ & CPU time${}^*$ \\
  \midrule
  $T_1\approx 0.36$ & $-9.445(77)$  &  $-2.545(45)$  & $704$ & $3.3$ Mch\\
  $T_2\approx 4$    & $-11.426(65)$ &  $-4.700(80)$  & $548$ & $0.6$ Mch\\
  $T_3\approx 45$   & $-11.994(57)$ &  $-5.200(88)$  & $654$ & $0.5$ Mch\\
  \bottomrule
  \end{tabular}
  \begin{tablenotes}
  \footnotesize 
  \item $^*$ Simulations performed at Marconi (CINECA) using 4096
  cores of Intel Xeon Phi 7250 Knights Landing (KNL) processors.
  \end{tablenotes}
  \end{threeparttable}  
  \caption{Results for the bare expectation values $\langle T^{F,G}_{0k} \rangle_\xi$
  at the three chosen temperatures: $T_1$, $T_2$, $T_3$. A total of $N_{\rm ms}$ 
  measures was considered, each separated by 10 MDU. The estimated CPU time required
  for the generation of the ensembles is also given.}
  \label{tab:Results}
\end{table}

The results for the  bare expectation values $\langle T^{F,G}_{0k} \rangle_\xi$
corresponding to the ensembles $T_1,T_2,T_3$ are given in Table \ref{tab:Results}. 
The table provides the total 
number of measurements gathered. These were collected every 10 MDU along the Monte 
Carlo histories, detecting no autocorrelations. In the table we also report 
the CPU time invested in the calculations. As one can see, the figures are modest,
while the statistical precision reached on the bare quantities is remarkable: 
$0.5-1\%$, depending on the temperature, for the fermionic contribution
$\langle T^{F}_{0k} \rangle_\xi$, and slightly below $2\%$ for the gluonic
contribution $\langle T^{G}_{0k} \rangle_\xi$. The results are particularly
encouraging since, as expected for $\Nf=3$ QCD, the fermionic contribution
to the entropy is significantly larger than the gluonic one, but it can be 
determined more accurately for a given number of independent measurements,
especially at high temperature.

A few miscellaneous observations are in order at this point. Firstly, if we had
measured the observables more frequently, we would have obtained a significantly
better precision at essentially the same computational cost. At $T_2$ and $T_3$
for example, we observed that the autocorrelations of $T^{F,G}_{0k}$ are in fact 
below 2 MDU: much shorter than we originally expected. As the generation 
of a gauge configuration is for all ensembles substantially more expensive than
the measurement of the observables (about a factor 9 larger), measuring every 2 
MDU instead of 10 MDU would reduce the computational effort necessary to obtain 
a given statistical precision by an interesting factor. Secondly, for these 
exploratory runs we did not consider any optimization of our code to run on KNL 
processors. Another interesting speed-up factor might thus be obtained even
with a basic optimization strategy.%
	\footnote{By comparing timings obtained on the KNL processors
	at Marconi with those obtained on our local cluster of 
	Intel Xeon E5-2630 and Intel Xeon E5-2650 processors, we estimated
	a loss in performances by a factor $2-3$.}

When moving from intermediate to high temperatures we observe that a precise
determination of the bare expectation values $\langle T^{F,G}_{0k} \rangle_\xi$
becomes easier. The main reasons is because the simulations do become easier. 
This is so because the theory to be simulated approaches the simple free theory, 
and the spectral gap of the Dirac operator gets larger as the temperature
increases; this is a similar situation to step-scaling simulations with the
SF~\cite{Brida:2016flw,DallaBrida:2016kgh}. In addition, we note that the 
relative error of the fermionic contribution $\langle T^{F}_{0k} \rangle_\xi$
decreases for a given number of independent measurements.

Finally, it is also interesting to note that the simulation algorithm appears 
to sample topology well at the lowest temperature $T_1$, where 
$a\approx 0.064\,\fm$. There we find: $\langle Q^2\rangle_{\xi}=5.2(3)$.%
  \footnote{We monitored topology  by means of the topological charge 
  $Q$ defined through the Yang-Mills gradient flow~\cite{Luscher:2010iy}. 
  In particular, we measure the topological charge at flow-time 
  $\sqrt{8t}=0.4\times T^{-1}$, and \emph{define} the ``$Q=0$ sector''
  as the set of gauge configurations for which $|Q|<0.5$.}
At the higher temperatures $T_2$ and $T_3$, however, quantitative conclusions
cannot be drawn. Certainly, one would expect topology to be very much suppressed
at these high temperatures, but concurrently the simulation algorithm must suffer
from a severe freezing of the topology at these very small lattice 
spacings~\cite{DelDebbio:2002xa,Schaefer:2010hu}. All we can say is that our runs
seem to exclusively sample the $Q=0$ sector.

\section{Conclusions}

The results we presented show that within the framework of 
SBC, a precision of $0.5-1\%$ on the bare quantities entering the determination
of the entropy density is reachable with modest computational effort and
over a wide range of temperatures, easily covering two orders of magnitude
in the temperature. This is very encouraging in view of a precise 
determination of the EoS of $\Nf=2+1$ QCD using non-perturbatively O($a$)-improved 
Wilson fermions. In particular, differently from the situation that occurs with
more standard methods, the higher the temperature is, the easier the simulations
become, and thus the easier is to obtain precise results. The fundamental reason
for this favorable situation is that the problem of computing the bare expectation
values and their renormalization are completely disentangled in our framework, and 
can therefore be tackled separately. The important question at this
point is hence whether the necessary renormalization of the EMT can be obtained with
a competitive level of precision. This issue is under current investigation and the
details of our renormalization strategy are in preparation~\cite{Draft:2017}. 

Regarding the renormalization of the EMT we must note that several interesting 
ideas, based on the Yang-Mills gradient flow~\cite{Luscher:2010iy}, have been 
proposed~\cite{Suzuki:2013gza,Makino:2014taa,DelDebbio:2013zaa} (see~\cite{Suzuki:2016ytc}
for a review). Following some of these strategies promising results for the EoS
have already been obtained, both in $\SUthree$ Yang-Mills theory and 
QCD~\cite{Taniguchi:2016ofw,Kitazawa:2016dsl}. From a different perspective, other
promising ideas for determining the EoS have been devised and 
tested~\cite{Caselle:2016wsw,Nada:2017xmz}. 

Finally, another important issue we shall address in order to obtain an accurate 
determination of the EoS using Wilson fermions is a systematic study of the 
O($a$)-improvement of the EMT. In this respect, it is interesting to note that 
chiral symmetry ``restoration'' at high temperature may have interesting consequences
on the lattice artifacts of the theory. Following standard arguments~\cite{Frezzotti:2003ni}, 
one expects indeed some degree of  ``automatic'' O($a$)-improvement at high temperature~\cite{Draft:2017}.

\section{Acknowledgements}

The authors thank Bastian Brandt for discussion on the development of the code. 
M.D.B. would like to express his gratitude to the Theoretical Physics Department
at CERN for the hospitality extended to him, and also thanks Stefan Sint for 
interesting discussions. The code used for the generation of the ensembles is 
based on the \texttt{openQCD-1.6} package~\cite{LuscherWeb:2016}.
Numerical calculations have been made possible through a CINECA-INFN agreement, 
providing access to resources on GALILEO and MARCONI at CINECA. The authors are 
also grateful to the computer centers at the University of Milano-Bicocca 
and CERN for their valuable support and resources.

\bibliography{lattice2017}

\end{document}